\documentclass[10pt,twoside,a4paper]{article}
\usepackage[dvips]{epsfig}

\newcounter{eqAnsatzPointInteraction}

\newcounter{eqAlternativeStatement}

\newcounter{eqInteractionTilde}

\begin{document}

%

\title{
{\bf The Beginning of Chiral Symmetry}
\author{G.~Kramer$^1$, H.~Rollnik$^2$ and B.~Stech$^3$
\vspace{2mm} \\
{\normalsize $^1$ II. Institut f\"ur Theoretische
  Physik, Universit\"at Hamburg} \vspace{2mm}\\ 
\normalsize{$^2$ Physikalisches Institut,
  Universit\"at Bonn} \vspace{2mm}\\ 
\normalsize{$^3$ Institut f\"ur Theoretische Physik, Universit\"at
 Heidelberg} \vspace{2mm}\\ }}
\maketitle
\begin{abstract}
\noindent
For readers interested in the history of chiral symmetry we present the 
translation of two papers from 1955 and 1959 from German to English, in
which chiral symmetry properties for weak and strong interactions have been
postulated and discussed.
\end{abstract}
\clearpage

Chiral symmetry is one of the fundamental properties of the theory of particle physics. It requires invariance of the Lagrangian under independent global transformations of left and right handed fermion fields. For a single particle generation, the standard model Lagrangian of particle physics, with its electroweak and  strong gauge and Yukawa interactions, has this symmetry. This holds also in natural extensions of the standard model, the grand unified theories, such as $SO(10)$ and $E6$. For several generations the Yukawa interaction is no more invariant, but chiral symmetry can be restored, if a generation symmetry is built into the model.  The fermion masses then arise from the spontaneous breaking of chiral and generation symmetry. 
 
For some readers it may therefore be of interest to look at the very first suggestions of chiral symmetry published many years ago. For their convenience we translate here from German to English the two original papers dealing with chiral symmetry for weak and strong interactions, published in Zeitschrift 
f\"ur Physik in 1955 \footnote{B.~Stech and J. H. D. Jensen, Z. Phys. 141, 175
(1955)} and 1959 \footnote{G. Kramer, H. Rollnik and B. Stech, Z. Phys. 154, 
564 (1959)} respectively. The fact that they have been written in German may 
be the reason why they are less known than later papers on this subject. 
In the translation we tried to be as close as possible to the originals.
In the first paper, chiral symmetry  of the weak interaction Lagrangian was postulated in the form of a $\gamma_5$ symmetry. Because this paper was written before Lee and Yang suggested the existence of parity changing processes, the symmetry operation consisted of the simultaneous replacement of two fermion field operators by $\gamma_5$ times these fields. After the discovery of parity changing transitions, the extension of the symmetry by simply transforming each fermion field separately with $\gamma_5$ became possible. The corresponding invariance requirement led immediately to the unique and well-known  
chirality invariant form of the weak interaction.
 
The suggestion that chiral symmetry could also be a symmetry of the strong interaction, even though this interaction is parity conserving,
was made in the second paper we translate here. To our knowledge it was the first paper proving that chiral invariance could also be a property of the  strong interaction Lagrangian. A non-linear square root type meson-nucleon coupling served for this purpose. Today, one uses for the effective interaction with the pion field a more convenient expression with the meson field in the exponent. In both papers presented here chiral symmetry is valid for the interactions, but is still violated by the mass terms in the free part of the Lagrangian. It is a symmetry "as if the masses were zero". Only later  this deficiency could be removed with an improved form of the effective Lagrangian for mesons and baryons and, for the basic theory,  by generating the fermion masses using the Higgs mechanism.

\cleardoublepage
\setcounter{footnote}{0}
\setcounter{equation}{0}
\setcounter{page}{1}


\begin{center}
{\large\bf The coupling constants in the theory of $\beta$ decay\footnote{To the $60^{\mbox{th}}$ birthday of Professor Dr. {\sc H. Kopfermann}.}.}\\
\vspace{0.5cm}
By\\
\vspace{0.2cm}
{\sc Berthold Stech} and {\sc J. Hans D. Jensen}.\\
\vspace{0.2cm}
{\em (Received on February $15^{\mbox{\footnotesize th}}$, 1955)}\\
\end{center}

{\footnotesize By only using data about $\beta$ decay (not $\mu$ meson decay) the restrictions in the choice of the coupling constants 
 well secured by experiment are discussed. After that two ``a priori'' arguments will be described where one of them yields the alternative:
($STP$) or ($VA$). The other ``a priori'' argument determines the exact value of the coefficients, namely: $S-T+P$ and $V-A$ respectively.}

\begin{center}
{\em Introduction.}
\end{center}
The theory of $\beta$ decay is still suffering from the ambiguity of the possible ans\"atze for the interaction operator, even if one restricts
oneself to a  point interaction of the four fermions participating in the process. The most obvious ansatz would be a simple scalar one for the interaction operator

\begin{equation}
W = U + U^{*} \quad \mbox{with} \quad U=g \cdot S \label{interaction}
\end{equation}
where $S$ stands for the abbreviation 

\begin{equation}
S = (\bar{\psi}_{p} \cdot \psi_{n})(\bar{\varphi}_{e} \cdot \varphi_{\nu}).
\end{equation}
Here $g$ is a constant measuring the coupling of the four fermion fields; 
 $\bar{\psi}=\psi^{\dagger}\gamma_{4}$, where $\psi^{\dagger}$
is the hermitean conjugate spinor and $\psi_{p}, \psi_{n}, \psi_{e}, \psi_{\nu}$ are the field functions (operators) of  proton, neutron,
electron and neutrino, respectively. Such an ansatz is not sufficient to describe all $\beta$ processes since this interaction ansatz implies that
the emerging proton is of the same spinor type as the disappearing neutron. Thus, the spin of the nucleon could not flip
in the $\beta$ process. However, the experimental data reveal that $\beta$ processes with spin flip occur with approximately the same
probability as those in which the nucleon keeps its spin orientation\footnote{For a discussion of the experimental material
see the report by {\sc Konopinski} and {\sc Langer} [Ann. Rev. of Nucl. Sci. {\bf 2}, 261 (1953)] and the literature cited there, and also
{\sc J. B. Gerhart}: Phys. Rev. {\bf 95}, 288 (1954); {\sc B. M. Rustad} and {\sc S. L. Ruby}: Phys. Rev. {\bf 89}, 880 (1953);
{\sc J. S. Allen} and {\sc W. K. Jentschke}: Phys. Rev. {\bf 89}, 902 (1953) and {\sc D. C. Peaslee}: Phys. Rev. {\bf 91}, 1447 (1953).}.
Therefore it is necessary to take into account the possibility of such spin flip processes in the ansatz for the interaction.
For allowed transitions and in a non-relativistic description of the nucleons, this happens in the generalization suggested by {\sc Gamow}
and {\sc Teller}:

\begin{equation}
U = g_{F}S+g_{G}G
\end{equation}
with $S$ as above, and $G=(\bar{\psi}_{p} \vec{\sigma} \psi_{n})(\bar{\varphi}_{e} \vec{\sigma} \varphi_{\nu})$ where
$\vec{\sigma}$ is {\sc Pauli}'s resp. {\sc Dirac}'s spin vector operator.

{\footnotesize A somewhat simpler description of this case is obtained using an ansatz for the interaction in which in each scalar spinor product
a heavy and a light particle are grouped together, e. g.
\begin{equation}
(\bar{\psi}_{p} \cdot \varphi_{\nu})(\bar{\varphi}_{e} \cdot \psi_{n}). \label{light_heavy_spinor}
\end{equation}
Here the spin of the emerging nucleon is not coupled any more to the spin of the disappearing nucleon, and already a simple scalar
ansatz of the type (\ref{light_heavy_spinor}) contains both {\sc Fermi} and {\sc Gamow-Teller} transitions. We will come back to this
later in \S 3. For the time being we shall use the traditional teminology.
}

For the general discussion including ``forbidden'' transitions it is necessary to treat also the nucleons relativistically since
relativistic terms of the nucleons contribute to the ``forbidden'' transitions to the same order of magnitude as the finite ratio
of the spatial extension of the nucleon fields to the wave lengths of the electron and neutrino states.
The most general relativistic invariant ansatz for a point interaction is, cf. {\sc Konopinski}, 1.c. -- (in the following we will write for
the field functions of the particles simply the particle symbol : $n$ for $\psi_{n}$  and accordingly $p$ for $\psi_{p}$,
$e$ for $\psi_{e}$, and $\nu$ for $\psi_{\nu}$) -- :

\begin{equation}
U = g_{S}S+g_{V}V+g_{T}T+g_{A}A+g_{P}P \label{int_expr}
\end{equation}

with

\[
\left.
\begin{array}{lll}
S &=& (\bar{p}n)(\bar{e}\nu) \\
V &=& \sum\limits_{\lambda}(\bar{p} \gamma_{\lambda} n)(\bar{e} \gamma_{\lambda} \nu)\\
T &=& \sum\limits_{\lambda ' < \lambda}(\bar{p} i\gamma_{\lambda '}\gamma_{\lambda } n)(\bar{e} i\gamma_{\lambda '} \gamma_{\lambda}\nu)\\
A &=& \sum\limits_{\lambda '' < \lambda ' < \lambda}(\bar{p} i\gamma_{\lambda ''}\gamma_{\lambda '}\gamma_{\lambda } n)(\bar{e} i\gamma_{\lambda ''}\gamma_{\lambda '} \gamma_{\lambda}\nu)=\sum\limits_{\lambda}(\bar{p} i\gamma_{5}\gamma_{\lambda } n)(\bar{e} i\gamma_{5} \gamma_{\lambda}\nu)\\
P &=& (\bar{p} \gamma_{5} n)(\bar{e} \gamma_{5} \nu)
\end{array}
\right\} (\theequation \mbox{a})
\]
\setcounter{eqAnsatzPointInteraction}{\theequation}

Here $\gamma_{\lambda}$ are the hermitean {\sc Dirac} matrices with the commutation relations
\begin{equation}
\gamma_{\lambda }\gamma_{\lambda '}+\gamma_{\lambda '}\gamma_{\lambda } = 2\delta_{\lambda '\lambda} \quad \mbox{for} \quad \lambda=1,2,3,4.
\end{equation}
For the product $\gamma_{1}\gamma_{2}\gamma_{3}\gamma_{4}$ we write as usual $\gamma_{5}$. The single round brackets transform
as a scalar in $S$, as a four vector in $V$, as an antisymmetric tensor in $T$, as a pseudo vector in ${A}$ and as a pseudo scalar in $P$.

The fundamental problem of the theory of $\beta$ decay is the finding of experimental rules or of theoretical arguments restricting the choice
of the five coupling constants $g_{S},g_{V},g_{T},g_{A},g_{P}$ and determining them unambiguously. 

{\footnotesize The reality of the coupling constants is guaranteed by the requirement that the interaction ansatz should be invariant
by simultaneously rewriting all four fermions into their anti particles\footnote{Cf. e. g. {\sc De Groot, H. A.}, and {\sc S. R. Tolhoek}
Physica, Haag {\bf 16}, 456 (1950). --- Phys. Rev. {\bf 84}, 150 (1951). --- {\sc Biedenharn, L. C.}, and {\sc M. E. Rose}: Phys. Rev.
{\bf 83}, 459 (1951).}
\footnote{Cf. also {\sc Pais} and {\sc Jost}: Phys. Rev. {\bf 87}, 871 (1952).}.
}

The theoretical ``a priori principles'' previously suggested to give  restrictions for the choice of $g's$ --- (they are discussed
in {\sc Konopinski} and {\sc Langer} l.c.
elaborately)
---, have lead to results that contradict the experimental results.

{\footnotesize
{\sc Peaslee}\footnote{{\sc Peaslee}, 1. c., cf. footnote 8 on the next page.}  gives an analysis of the experimental results, according to which a combination  $S-T+P$, i. e.
$g_{S}=-g_{T}=g_{P}$ and $g_{V}=g_{A}=0$, is not in contradiction with any experimental result. He then discusses a theoretical
selection principle which is in accordance with this combination. However his principle is based on a special relation\footnote{
Cf. also {\sc E. J. Konopinski} and {\sc H. M. Mahmud}: On the Universal Fermi Interaction, Phys. Rev. {\bf 92}, 1045 (1953).
We thank Mr. {\sc Konopinski} for sending the 
manuscript before publication} 
between $\beta$ processes and $\mu$ meson decay which fixes the spectrum of the decay electrons of the $\mu$ meson
-- (the spectrum goes to zero at the maximum electron energy). But according to the most precise measurements\footnote{
{\sc Bramson, H., A. Seifert} and {\sc W. Heavens}: Phys. Rev. {\bf 88}, 304 (1952). -- {\sc Sagane, R.} and collaborators: Phys. Rev.
{\bf 95}, 863 (1954). -- {\sc Harrison, F. B.} and collaborators: Phys. Rev. {\bf 96}, 1159 (1954). --
{\sc Vilain, J. H.} and {\sc W. Williams}: Phys. Rev. {\bf 94}, 1011 (1954).}
the spectrum has a finite value at the energy boundary.}

In the present article we like to summarize the consequences for the interaction ansatz resulting from experiments
 restricted to $\beta$ processes and to search for  and to discuss the simplest ``theoretical'' principles which are in accordance with
the empirical results and tighten them.


\begin{center}
{\em \S1. A simple restriction for the coefficients in the interaction ansatz}
\end{center}
The alternative statement with the best experimental confirmation reads, cf. {\sc Konopinski} and {\sc Langer}, 1. c.\\

{\em either} one has\\
\stepcounter{equation}
\setcounter{eqAlternativeStatement}{\theequation}
\parbox{11cm}
{
\[
 g_{V} = g_{A} = 0
 \]
} \hfill
\parbox{6mm}
{
\[
( \theequation \mbox{a})
\]
}

{\em or}\\
\parbox{11.5cm}
{
\[
 g_{S} = g_{T} = g_{P} = 0.
 \]
} \hfill
\parbox{6mm}
{
\[
( \theequation \mbox{b})
\]
}

In other words: there can either be a $(STP)$ combination or a $(VA)$ combination\footnote{\label{notation_fnote}The notation $(AV)$ etc. means that at least
one of the coefficients $g$ of the terms in brackets are not equal to zero while the $g$ factors of the terms not appearing in the brackets
vanish.}. This statement is affirmed by precise measurements of electron and positron spectra of allowed and forbidden
transitions\footnote{This is equivalent with the absence of all so-called ``{\sc Fierz} terms'' in the spectrum, {\sc Konopinski} and {\sc Langer},
1.~c.}. Of course, the exact vanishing of the relevant groups of $g$ coefficients cannot be extracted from experiments.
However, one would certainly come to  a contradiction with the measured spectra if the absolute value of one of the $g$ factors having been set to zero
would be bigger than $1/10$ of the value of the biggest $g$ coefficient of those having not been set to zero. This is a careful
estimation, probably one can conclude that the corresponding $g$ values must be even closer to zero.

There exists a very simple theoretical argument that provides exactly the alternative expressed in (\theequation ).
First it may be formulated quite formally:

One requires that the interaction (\ref{int_expr}) -- except for a sign -- will transform into itself if one performs simultaneously
the replacement ($e$ and $\nu$ denote, as above, the field functions):


\stepcounter{equation}

\begin{equation}
e \rightarrow \gamma_{5} e \qquad \nu \rightarrow \gamma_{5} \nu. \label{transformation}
\end{equation}

One finds from a  quite elementary calculation that under the given replacement $V$ and $A$ transform into themselves while $S$, $T$ and $P$
change their sign:

\begin{eqnarray*}
(\overline{\gamma_{5} e} \gamma_{\lambda}  \gamma_{5} \nu ) &=& (\overline{e} \gamma_{\lambda} \nu ) \qquad \mbox{and accordingly for the terms in $A$,}\\
(\overline{\gamma_{5} e} \gamma_{5} \nu ) &=& -(\overline{e} \nu ) \qquad \mbox{and accordingly for the terms in $T$ and $P$.}
\end{eqnarray*}

The requirement of invariance under this transformation excludes therefore that $(VA)$ can appear together with $(STP)$ in the interaction
operator.

With respect to the physical interpretation of the transformation (\ref{transformation}) it may be pointed out that the replacement of
$\psi$ by $\gamma_{5}\psi$ is equivalent to the replacement of $m$ by $-m$ in the {\sc Dirac} equation, for from

\stepcounter{equation}

\parbox{10cm}
{
\[
\bigg( \sum_{\lambda} \gamma_{\lambda}\frac{\partial}{\partial x_{\lambda}} 
+ \frac{mc}{\hbar} \bigg) \psi= 0
 \]
} \hfill
\parbox{8mm}
{
\[
( \theequation \mbox{a})
\]
}

follows

\parbox{10cm}
{
\[
\bigg( \sum_{\lambda}  \gamma_{\lambda}\frac{\partial}{\partial x_{\lambda}} - \frac{mc}{\hbar} \bigg) \gamma_{5} \psi= 0.
 \]
} \hfill
\parbox{8mm}
{
\[
( \theequation \mbox{b})
\]
}

If the electron would have, like the neutrino, a vanishing rest mass then, with $e$ being a solution of the {\sc Dirac} equation,
$\gamma_{5}e$ would be a solution as well. The finite value of the rest mass of the electron  is probably irrelevant for the general form of the
interaction operator. Thus it can make good sense to require: ``{\em The interaction operator should have the same general form as
if the rest mass of the electron would be zero}.'' But then the invariance of the interaction operator under the transformation
(\ref{transformation}) is an almost compelling requirement\footnote{The invariance of the interaction operator under the
$\gamma_{5}$ transformation of the electron and neutrino states has the consequence that all terms linear in the electron mass
(``{\sc Fierz} terms'') in the spectrum vanish. In {\sc Peaslee} 1.~c. a $\gamma_{5}$ invariance for two undistinguishable neutrinos
in $\mu$ decay has been required. Only in this case the requirement of invariance including the sign suggests itself.}.

{\footnotesize
One still could tighten this selection principle in the following sense. If one presumes that the $\beta$ interaction operator
has the same form as for four different fermions which all have a vanishing rest mass, then one could require the invariance of the
interaction operator under the simultaneous replacement of every arbitrary pair of field functions $\psi$ by $\gamma_{5}\psi$;
so e.~g. also for $e \rightarrow \gamma_{5}e$ and $p \rightarrow \gamma_{5}p$.
With this strict requirement the only combinations possible would be:

\[
\begin{array}{lll}
\mbox{either}\,\, & g_{S} = g_{T} = g_{P} = 0; \quad g_{A}=\pm g_{V}, \quad &\mbox{i. e.:}\quad  V+A \quad\mbox{or}\quad V-A\\
\mbox{or} & g_{V} = g_{A} = 0; \quad g_{P}=+g_{S}; \quad g_{T} \,\,\mbox{arbitrary,}\quad &\mbox{i. e.:}\quad  S+\mbox{const}\, T + P\\
\mbox{or} & g_{V} = g_{A} = 0; \quad g_{P}=-g_{S}; \quad g_{T}=0, \quad &\mbox{i. e.:}\quad  S-P.
\end{array}
\]

We suppress the explicit proof that this selection follows from the invariance requirement mentioned, since it requires some
calculations, and also because in the next paragraph we will discuss an invariance requirement  more plausible to us having
consequences which  are in accord with the selections mentioned just now, but  tighten this selection even further.}

For the following we will only stress on the existence of the experimentally well established alternative
(\arabic{eqAlternativeStatement}) and the fact that one can obtain  it from a very little incisive theoretical requirement.


\begin{center}
{\em \S2. Fixing the coefficients in the interaction ansatz}
\end{center}
The statement (\arabic{eqAlternativeStatement}) is still very general since it states nothing about the relative magnitude of
the coefficients in the combination $(STP)$ and in $(VA)$ respectively. Before coming to this point we will first quote additional experimental
findings.

From the measurements of $f\cdot t$ values of some specific allowed transitions it can be concluded, cf. especially \footnote{{\sc Gerhart, J.~ B.}:
1.c.} and \footnote{{\sc Kofoed-Hansen, O.}: Phys. Rev. {\bf 92}, 1075 (1953). --- {\sc Staehelin, P.}: Phys. Rev. {\bf 92}, 1076 (1953). },
that $|g_{G}/g_{F}|$ is close to 1. This implies for the case of a $(STP)$ interaction that $|g_{T}/g_{S}|\approx 1$ and for the case
of a $(VA)$ interaction $|g_{A}/g_{V}|\approx 1$. On the other hand, the ratio $g_{P}/g_{S}$ is within a large range undetermined.

Now we would like to point out that the special combinations
\[
\begin{array}{lll}
\mbox{either}\,\, & g_{S} = -g_{T} = g_{P} \quad &\mbox{and} \quad g_{A}=g_{V}=0, \quad \mbox{i. e.:}\quad  S-T+P\\
\mbox{or} & g_{A} = -g_{V} \quad &\mbox{and} \quad g_{S}=g_{T}=g_{P}=0, \quad \mbox{i. e.:}\quad  A-V,
\end{array}
\]
follow from a very simple isotropy requirement. From the known general formulae obtained for the angular correlation of $\beta$-decay\footnote{Cf. e. g.
{\sc Groot, S. R. de,} and {\sc H. A. Tolhoek}: 1.c.} it follows that for $|g_{S}|=|g_{T}|$ and $g_{V}=g_{A}=0$ or for
$|g_{V}|=|g_{A}|$ and $g_{S}=g_{T}=g_{P}=0$ there is no angular correlation between electron and the neutrino in the
decay of the free neutron\footnote{The angular correlation in non-relativistic approximation is proportional to
\[
\frac{1}{3}|\bar{p}\vec{\sigma}n|^2(g_{T}^{2}-g_{A}^{2})-|(\bar{p}n)|^{2}(g_{S}^{2}-g_{V}^{2});
\]

but for the decay of the free neutron and the square of the {\sc Gamow-Teller} matrix element $(\bar{p}\sigma n)$ is three times the square of
the {\sc Fermi} matrix element $(pn)$.}.
This can be understood from a  plausible general isotropy requirement, if one describes the $\beta$ decay, as usual, by the annihilation of
two particles (neutron and neutrino) and the creation of two others (proton and electron)\footnote{In the representation
$n \rightarrow p+e+\nu^{C},~ \nu^{C}$ denotes an ``antineutrino'', cf. also {\sc Mahmoud} and {\sc Konopinski} 1.c.}.
In a reference frame, in which the neutron and the vanishing neutrino (which is in a state of negative energy) have opposite momenta
and thus total momentum zero, no direction is preferred and the following requirement seems plausible: {\em ``In this reference frame
the distribution of the direction of the emitted protons should be isotropic''.} Since the electrons have the opposite momenta of the protons in this
reference frame, also the distribution of the directions of the electrons will then be isotropic.


\begin{center}
{\em \S3. Consequences of the isotropy requirement}
\end{center}
One can overlook the consequences of this isotropy requirement best in a representation of the interaction operator differing from the
usual one, namely, in which in analogy with the expression 
(\ref{light_heavy_spinor}), each nucleon is coupled 
with one of the light particles. In this case the two annihilated, and the two created fermions as well, can be combined to form   {\sc Lorentz} covariant expressions. For this purpose it is necessary to introduce the so-called conjugate spinors, cf. e.~g.
{\sc Pais} and {\sc Jost} 1.~c. These are defined as follows: To every spinor $\psi$ a conjugate spinor $\psi^{C}$ is defined by the following relation:
\begin{equation}
\psi^{C} = C\bar{\psi}^{T}.
\end{equation}
The meaning of $\bar{\psi}$ is given above near eq. (\ref{interaction}), $T$ shall indicate the transposition, i.~e. the interchange of columns
and rows.
The unitary transformation matrix $C$ is fixed by the following properties\footnote{In the {\sc Dirac} representation of the $\gamma$ ´s
the matrix $C$ has, up to an arbitrary factor $\pm \sqrt{\pm1}$, the form
\[
C=
\left(
\begin{array}{rrrr}
0  &  0  &  0  &  1 \\
0  &  0  & -1  & 0 \\
0  & 1   &  0  & 0 \\
-1 & 0  & 0   & 0
\end{array}
\right)
\]
}

\[
C^{\dagger}C=1, \qquad C^{T}C^{-1}=-1, \qquad C^{-1}\gamma_{\lambda}C=-\gamma_{\lambda}^{T}
\]
$\psi^{C}$ and $\psi$ transform under {\sc Lorentz} transformations in the same way; they satisfy the  {\sc Dirac} equation
for free particles but differ by having eigenvalues with opposite signs for  energy, momentum and spin components.

With this notation the five invariants $S, V, T, A, P$ of eq. (\arabic{eqAnsatzPointInteraction}a) are linear combinations of the five  new invariants 

\begin{equation}
\left.
\begin{array}{lll}
\tilde{S} &=& (\overline{\nu^{C}}n)(\overline{p}e^{C}) \\
\tilde{V} &=& \sum\limits_{\lambda}(\overline{\nu^{C}} \gamma_{\lambda} n)(\overline{p} \gamma_{\lambda} e^{C})\\
\tilde{T} &=& \sum\limits_{\lambda ' < \lambda}(\overline{\nu^{C}} i\gamma_{\lambda '}\gamma_{\lambda } n)(\overline{p} i\gamma_{\lambda '} \gamma_{\lambda}e^{C})\\
\tilde{A} &=& \sum\limits_{\lambda}(\overline{\nu^{C}} i\gamma_{5}\gamma_{\lambda } n)(\overline{p} i\gamma_{5} \gamma_{\lambda}e^{C})\\
\tilde{P} &=& (\overline{\nu^{C}} \gamma_{5} n)(\overline{p} \gamma_{5} e^{C})
\end{array}
\right\}
\end{equation}
The linear relation between the invariants (\arabic{eqAnsatzPointInteraction}a) and (\theequation) is\footnote{A similar transformation has been performed by {\sc M. Fierz} [Z. Physik {\bf 104},
553 (1937)]. Cf. also {\sc Caianiello, E. R.}: Nuovo Cim. {\bf 8}, 534, 749 (1941); {\bf 9}, 336 (1952), and {\sc Michel, L.}:
Proc. Phys. Soc. A {\bf 63}, 514 (1950).}:

\begin{equation}\label{linear_relation_invariants}
\left.
\begin{array}{lll}
-4\tilde{S} &=& (S-T+P)-(V-A) \\
-2\tilde{V} &=& 2(S-P)+(V+A)\\
-2\tilde{T} &=& (3S+T+3P)\\
-2\tilde{A} &=& 2(S-P)-(V+A)\\
-4\tilde{P} &=& (S-T+P)+(V-A). 
\end{array}
\right\}
\end{equation}
The isotropy requirement discussed at the end of \S 2 now says that the interaction potential
\begin{equation}
U=a_{S}\tilde{S}+a_{V}\tilde{V}+a_{T}\tilde{T}+a_{A}\tilde{A}+a_{P}\tilde{P} \label{interaction_tilde}
\end{equation}
-- applied to a process in which neutron and neutrino have 
opposite momenta -- is invariant under an arbitrary rotation of the
electron and proton directions.

Thus, the replacement
\begin{equation}
\bar{p} \rightarrow \bar{p}\Lambda^{-1}, \qquad e^{C} \rightarrow \Lambda e^{C},
\end{equation}
where $\Lambda$ denotes the transformation matrix\footnote{See e.~g.: {\sc Pauli, W.}: Handbuch der Physik XXIV, 1, p. 221. If the rotation is given by the {\sc Lorentz}
transformation of the space coordinates $x_{\lambda}'=\sum\limits_{\mu}a_{\lambda\mu}x_{\mu}$, then the corresponding transformation matrix
$\Lambda$ of the spinors is determined by $\Lambda^{-1}\gamma_{\lambda}\Lambda=\sum\limits_{\mu}a_{\lambda\mu}\gamma_{\mu}$.} 
acting on
the spinor components should leave $U$ unchanged.
(The momentum conservation insures that the exponential functions of the plane waves do not have to be considered). Because of the known
transformation properties of the vector, tensor etc. combinations the  invariance holds  only  for
\begin{equation}
a_{V}=a_{T}=a_{A}=0 \label{selection_V_T_A_zero}
\end{equation}
e.i. only the ``scalar'' and ``pseudoscalar'' expressions $\tilde{S}$ and $\tilde{P}$ can contribute to the interaction.
From (\ref{linear_relation_invariants}) one has now the strict requirement that $U$ can only be a linear combination of
$-2(\tilde{S}+\tilde{P})=S-T+P$ and $2(\tilde{S}-\tilde{P})=V-A$. Taking the alternative (\arabic{eqAlternativeStatement}) into account then
\stepcounter{equation}
\setcounter{eqInteractionTilde}{\theequation}
\[
\begin{array}{llr}
\mbox{either}\,\, & U = -g(S-T+P) = 2g(\tilde{S}+\tilde{P}) & \qquad\qquad (\theequation\mbox{a}) \\
\mbox{or} & U = -g(V-A) = 2g(\tilde{S}-\tilde{P}) & \qquad\qquad (\theequation\mbox{b})
\end{array}
\]
are the only possible forms for the interaction. 

{\footnotesize
It shall be pointed out, that the alternative (\arabic{eqAlternativeStatement}) is only needed in the weak form $g_{S} g_{V} = g_{T} g_{A} = 0$ (i.~e. that $V$ 
cannot appear together with $S$
and that $A$ cannot appear together with $T$). This, however, is  well established experimentally, because it can already be concluded from
the precisely measured spectra of allowed transitions  \footnote{Cf. {\sc Konopinski} and {\sc Langer}: 1.~c.}.}

In the formulation of the isotropy principle a reference system was used, in which under rotations only the spinor components
are transformed. Therefore, one can interpret the ``isotropy'' as a decoupling of the spin orientations of the created particles from the spin
orientation of the annihilated particles. This point of view can also be formulated quantitatively: For free particles the {\sc Hamilton} operator
commutes\footnote{$\vec{p}=\frac{\hbar}{i}\vec{\nabla}$; in the representation of the states by plane waves the operator
$\vec{p}$ can be replaced by its eigen value.} 
with $(\vec{\sigma}\cdot \vec{p})$ and thus commutes for particles with a fixed  momentum direction 
with $(\vec{\sigma}\cdot \vec{e})$, if the unit vector $\vec{e}$ points in the  momentum direction.
This operator is the operator of an infinitesimal spin rotation around the momentum vector.
If one now requires the invariance of the interaction operator $U$ under such spin rotations of the created particles, keeping the spins
of the annihilated particles  fixed,
\begin{equation}
\bar{p} \rightarrow \bar{p}(\vec{\sigma} \vec{e}) \qquad \mbox{and} \qquad e \rightarrow -(\vec{\sigma} \vec{e})e,
\qquad \mbox{i. e.} \quad e^{C} \rightarrow (\vec{\sigma} \vec{e})e^{C},
\end{equation}
then only such relativistic invariants can contribute to the interaction ansatz (\ref{interaction_tilde}), which contains $\gamma$ combinations
commuting with $(\vec{\sigma} \vec{e})$.
Commuting with $(\vec{\sigma} \vec{e})$, however, are only $1$ and $\gamma_{5}$. Also in this way one finds the result
(\ref{selection_V_T_A_zero}).
The latter leads in non-relativistic approximation (for the heavy particles) to $|g_{F}|=|g_{G}|$ and reveals here again and clearly the
decoupling of neutron and proton spin: spin flip and non-spin flip processes have the same probability.

{\footnotesize
The result (\ref{selection_V_T_A_zero}) ($a_{V}=a_{T}=a_{A}=0$) which has been obtained with the help of the isotropy requirement
is also in accord with the requirement of invariance (up to the sign) of the part of the interaction operator containing
the created (or the annihilated) particles alone under the particle-antiparticle conjugation\footnote{
For an earlier application of the particle-antiparticle conjugation of two particles in $\beta$ decay that, however, has led to results not compatible with the experimental results, see {\sc S. R. de Groot} and {\sc H. A. Tolhoek}, 1.~c.}.
One finds using  the definition\footnote{Cf. e.~g.: {\sc Peaslee, D. C.}: 1.~c. and {\sc Pais} and {\sc Jost}, 1.~c.}
\begin{eqnarray*}
P^{C}( \overline{\psi}_{1} \Omega \psi_{2} ) &=&( \overline{\psi_{2}^{C}} \Omega \psi_{1}^{C} )\\
P^{C}(\overline{p}e^{C})&=&(\overline{e}p^{C})=-(\overline{p}e^{C})\\
P^{C}( \overline{p} \gamma_{5} e^{C} )& =&( \overline{e} \gamma_{5} p^{C} ) = -(\overline{p} \gamma_{5} e^{C}).
\end{eqnarray*}
}

However, since the corresponding requirements as well as the one for time reversal separately for created and annihilated particles seems 
not very plausible, we dispense from deriving our result (\ref{selection_V_T_A_zero}) from such conditions.

{\em Note added in proof.} Both interaction ans\"atze (\arabic{eqInteractionTilde}a) $(S-T+P)$ and
(\arabic{eqInteractionTilde}b) $(A-V)$ are antisymmetric by interchanging  either the created particles among
themselves or the annihilated particles among themselves  as a simple calculation
shows, cf. also {\sc Peaslee} 1.~c., {\sc Caianiello} 1.~c. and {\sc Pursey}\footnote{{\sc Pursey, D. L.}: Physica, Haag {\bf 18}, 1017
(1952). In $\beta$ decay the pair of disappearing particles (``Paar der verschwindenden Teilchen'') is  identical with the pair of 
uncharged particles (``Paar der ungeladenen Teilchen'') $n, \nu$
for which {\sc Pursey} discusses an universal antisymmetry requirement.}.
This restricted part of the general antisymmetry requirement discussed by {\sc Critchfield} and {\sc Wigner}\footnote{
{\sc Critchfield, C. L.}: Phys. Rev. {\bf 63}, 417 (1943).} 
is therefore compatible
with our isotropy requirement discussed above. The isotropy requirement, however, is more restrictive.
The requirement of antisymmetry under the interchange  of  the created or the disappearing particles would also allow 
the combination\footnote{{\sc Nordheim, L.}: Report of the Indian Conference on Nuclear Spectroscopy, Indiana Univ. 1953.} $S-A-P$
which is not compatible with the isotropy requirement and the experimental findings.

The antisymmetry requirement for the ``created particles'' {\em together with} the invariance under the substitution (\ref{transformation}) is, however,
equivalent with the isotropy requirement {\em together with} the invariance postulate (\ref{transformation}).


\begin{center}
{\em \S4. The $S-T+P$ interaction}
\end{center}
A theoretical decision in favor of or against one of the two possibilities presented in (\arabic{eqInteractionTilde}), differing in the new representation only by a sign, has not yet
been accomplished. Thus, at this place only a discussion of the experimental results is possible.
Especially important to us are the recoil measurements  in $\beta$ decay of He$^{6}$ that have been performed
by {\sc Rustard} and {\sc Ruby}, 1.~c. and by {\sc Allen} and {\sc Jentschke}, 1.~c.
These measurements  show that $T$ is an essential part of the $\beta$ interaction operator.
They exclude the combination (\arabic{eqInteractionTilde}b). This finding is in agreement  with earlier evidences by {\sc Nordheim}\footnote{Cf.
e.~g.: {\sc Peaslee, D. C.}: 1.~c. und {\sc Pais} and {\sc Jost}, 1.~c.} and {\sc Peaslee}, 1.~c. that $P$ is likely a part of the interaction.

If one accepts this alternative and the theoretical selection principles  put forward, then $\beta$ decay is a $S-T+P$ interaction which,
in a new notation, takes the simple form
\begin{equation}
U = 2g(\tilde{S} + \tilde{P}) = 2g \{ (\overline{\nu^{C}}n) (\overline{p}e^{C}) +
(\overline{\nu^{C}} \gamma_{5} n) (\overline{p} \gamma_{5} e^{C}) \}. \label{ansatz_simple_form}
\end{equation}
One can see  that the tensor product $T$ appears in the interaction expression quasi only by an ``inapt choice of coordinates''.
To summarize: by taking  ``created'' and ``disappearing'' particles separately 
in one spinor product, only scalar and pseudoscalar ``interactions''
appear.
Apart from the  theoretical aesthetic point of simplicity  recommending the ansatz (\ref{ansatz_simple_form}),
the usage of this simple form means also a calculational simplification in quantitative computations of complicated -- e.~g.
multiply forbidden -- transitions, of the exact consideration of {\sc Coulomb} corrections etc. We hope to be able to present this in a later article.

\vspace{0.7cm}

{\em Heidelberg,} Institut f\"ur Theoretische Physik.

\cleardoublepage
\setcounter{footnote}{0}
\setcounter{equation}{0}
\setcounter{page}{1}



\renewcommand{\thefootnote}{\fnsymbol{footnote}}
\begin{center}
{\Large\bf
$\gamma _5$-Invariance and Parity Conservation in Strong Interactions}

\vspace{10mm}

{\large G.~Kramer, H.~Rollnik\footnote{On leave from Freie Universit\"at Berlin} 
and B.~Stech}

\vspace{2mm}
{Institut f\"ur Theoretische Physik, Universit\"at Heidelberg}

\vspace{2mm}

{\small (\it Received March 3 rd, 1959)}
\end{center}
\renewcommand{\thefootnote}{\arabic{footnote}}
\setcounter{footnote}{0}

\begin{abstract}
\medskip
\noindent
In this paper it is shown, that it is possible to construct parity conserving
baryon-pion interactions, which are $\gamma _5$ invariant in exactly the same
manner as electromagnetic and weak interactions.
\\
\end{abstract}

\clearpage

In previous publications \cite{1,2} the possibility was pointed out to 
construct couplings between baryons and $\pi$-mesons, which are invariant 
against space reflections and are in addition explicitly $\gamma _5$-invariant
in the same way as the weak and electromagnetic interactions. Here we 
understood as $\gamma _5$ invariance the invariance of the interaction
operator under the substitution $\psi _K \rightarrow \gamma _5 \psi _K$ for
any spin particle $K$, which appears in the coupling. For example,
neutron and proton are considered as different particles. Just this kind
of $\gamma _5$-invariance seems to be present in the weak interactions
\cite{3,4,5,6}. The proof in \cite{1,2} for the space-reflection invariance 
of certain $\gamma _5$ invariant $\pi$-meson-baryon couplings is not correct
\footnote[1]{We thank Dr. Symanzik and Dr. W. Theis for a hint in this 
respect}. In this paper a rigorous proof shall be given and the conditions 
explained, under which an explicitly $\gamma _5$-invariant $\pi$-meson-baryon 
interaction is simultaneously parity-invariant.\\

At first, for simplicity we consider only the $\pi$-meson-nucleon interaction.
The $\gamma _5$-invariant interaction part of the Lagrange density is of the 
following form
\begin{equation}
 L_W = \frac{f}{m} \bar{\psi}i\gamma _{\mu}(1+\gamma_5)\vec{T}\psi
\frac{\partial}{\partial x_{\mu}} \vec{\Phi}.  
\end{equation}
In this equation $\vec{T}$ stands for an isospin-vector matrix, which in 
general depends still on the $\pi$-meson field. Now, we try to determine
this isospin matrix $\vec{T}$ such, that the coupling (1) is space-reflection
invariant. Since the reflection invariance of a $\gamma _5$-invariant form
of coupling, as it is present in (1), is not easy to see, transformations
of the field operators should be performed. If we succeed to determine 
$\vec{T}$ in such a way, that a transformation of the field operators can be
given, which on one side leads to an obvious parity invariant coupling and on 
the other side does not change the physical consequences of the theory 
(S- matrix), then we have reached our goal.\\

A simple example is the coupling of neutral mesons, where $T$ naturally has 
only one component. In this case (1) is already with $T=1$ parity invariant.
Here, the simple substitution (directly in the Lagrange-function or in the 
field equations)
\begin{equation}
 \psi(x)=c \cdot e^{i\frac{f}{m}\Phi(x)} \psi '(x),~~\Phi(x)=\Phi '(x)
\end{equation}
leads to field equations in $\psi '(x)$ and $\Phi '(x)$ with a pure axial
coupling \cite{7}, $c$ is a renormalization constant (see Appendix). The
possibility, to eliminate the vector current $\bar{\psi}\gamma_{\mu}\psi$ 
from the field equations, follows from the vanishing of the divergence of 
this current. Whereas in the $\gamma _5$-invariant form of the Lagrange 
density and in the equations of motion the parity operator has a complicated 
form, for the primed field operators the usual form can be used and the 
reflection invariance can be seen directly. The physical equivalence of the 
transformed and untransformed field equations follows from the fact, that the 
ingoing and outgoing asymptotic fields of $\psi '(x)$, $\Phi '(x)$ agree with 
the corresponding fields of $\psi (x)$, $\Phi (x)$ and therefore lead to the 
same S-matrix. The proof of this assertion - also for the following more 
general case - will be given in the appendix.\\

\newpage
In the symmetrical meson theory the situation is more complicated, because of 
the presence of the non-commuting isospin matrices. In this case we use instead
of (2)  the following substitution \footnote[2]{In a somewhat different 
formulation the transformation (3) has been used several times for the proof of
equivalence theorems in the meson theory \cite{8,9,10,11}.}
\begin{equation}
 \psi (x) = c \cdot U\psi '(x),~~~ \Phi(x) = \Phi '(x).
\end{equation}
Here $U$ should represent a matrix invariant under rotations in isospace
of the following form
\begin{equation}
 U = \frac{1+i \vec{\tau} \vec{\Phi} w}{\sqrt{1+\vec{\Phi}^2 w^2}} , 
\end{equation}
where for the present $w$ is an arbitrary function of $\vec{\Phi}^2$ and 
$\vec{\tau}$ denote the usual isospin matrices. In order that the ingoing
and outgoing fields for $\psi '(x)$ and $\Phi '(x)$ agree with those for 
$\psi (x)$ and $\Phi (x)$, we require that $U$ allows a power series expansion
in $f/m$.\\

Performing the substitution (3) in the Lagrange operator or in the field 
equations the following new coupling arises:
\begin{equation}
L'_W =|c|^2 \bar{\psi}' i\gamma_ {\mu}  \Big\{ iU^{\dag} 
\frac{\partial }{\partial \vec{\Phi}'}U +\frac{f}{m} U^{\dag}
\vec{T}U (1+\gamma _5) \Big\} \psi'  \frac{\partial }{\partial x_{\mu}}
\vec{\Phi}' 
\end{equation}
The first term of this expression results from the interaction-free part 
of the Lagrange function.\\

Eq.(5) is certainly a reflection-invariant coupling, if the $\gamma _5$-free 
term contains even and the term multiplied with $\gamma_5$ contains odd powers
of the field $\Phi '(x)$. (This requirement implies that the meson field 
$\Phi '(x)$ is a pseudoscalar field.) Choosing in Eq.(1)
\begin{equation}
  \vec{T} = \frac{1}{2} \frac{1}{1+(f/m)^2 \vec{\Phi}^2}
   \left(\vec{\tau} + \frac{f}{m}[\vec{\Phi}\times\vec{\tau}] \right)
\end{equation}
the requirements are fulfilled, if for the transformation (3) and (4) the
function
\begin{equation}
 w(\vec{\Phi}^2) = \frac{f/m}{1+ \sqrt{1+(f/m)^2 \vec{\Phi}^2}}
\end{equation}
is used. $w(\vec{\Phi}^2)$ satisfies the requirement on the 
transformation $U$ made in connection with the asymptotic condition, namely 
the possibility of an expansion of it in powers of $f/m$.\\

Eq.(1) with (6) can be written in the following form:
\begin{equation}
L_W = \frac{f}{m}(\vec{j_{\mu}}+\vec{j^A_{\mu}})\left(
\frac{\partial}{\partial x_{\mu}} \vec{\Phi} -\frac{f^2}{m^2}
\frac{\vec{\Phi}}{1+(f/m)^2\vec{\Phi}^2} 
(\vec{\Phi}\frac{\partial}{\partial x_{\mu}}\vec{\Phi}) \right).
\end{equation} 
In this equation $\vec{j_{\mu}}$ denotes the total isovector current
following from (1) with (6), which obeys a continuity equation, whereas
$\vec{j^A_{\mu}}$ stands for the axial vector current
\begin{equation}
  \vec{j^A_{\mu}} = \bar{\psi} i \gamma _{\mu} \gamma _5 \vec{\tau} \psi .
\end{equation}
This kind of presentation of $L_W$ corresponds to the form aimed at in 
\cite{1,2}. The equation (4) in \cite{2} differs from (8) in the last term,
which however is necessary for the space-reflection invariance.\\

Written in the tranformed field the coupling has the following form, because of
$L(\psi)=L'(\psi ')$:
\begin{equation}
\left.
\begin{array}{l}
L_W'=|c|^2\ \Big\{ \frac{f}{m}\frac{1}{2}\bar{\psi}' i\gamma_{\mu}\gamma_5 
\vec{\tau}\psi' \Big(\frac{1}{a}\frac{\partial}{\partial x_{\mu}}\vec{\Phi}'
-\frac{f^2/m^2}{a^2(1+a)}\vec{\Phi}'(\vec{\Phi}' 
\frac{\partial}{\partial x_{\mu}}\vec{\Phi}')\Big) 
\\
\\
\phantom{L_W'=|c|^2\ \Big\{ \frac{f}{m}\frac{1}{2}\bar{\psi}' i\gamma_{\mu}\gamma_5 
\vec{\tau}}
-\frac{(f/m)^2}{a(1+a)}\frac{1}{2}\bar{\psi}'i\gamma_{\mu} \vec{\tau} \psi'
[\vec{\Phi}' \times \frac{\partial}{\partial x_{\mu}}\vec{\Phi}'] \Big\},
\end{array}
\right\}
\end{equation}
where
\begin{eqnarray}
\nonumber
  a= \sqrt{1+(f/m)^2 \vec{\Phi}'^2}
\end{eqnarray}
In Eq.(10) the reflection invariance is obvious, whereas the $\gamma _5$-
invariance for every particle, in contrast to (1), does not appear explicitly
anymore.\\

Of course, the form of Eq.(10) for $L_W'$ is only an example and is determined
by the specially chosen ansatz for $\vec{T}$ in Eq.(6). This $\vec{T}$ seems
to be the simplest one, which makes the coupling $\gamma _5$-invariant and
reflection invariant simultaneously.\\

The coupling to the electromagnetic field produces no new difficulties. It
turns out that the gauge invariant coupling of the photon to the Lagrange
function containing the interaction Eq.(1) leads to the same result as the 
explicitly reflection-invariant coupling to the Lagrange function with the
interaction Eq.(5) \footnote[3]{Because of the required $\gamma _5$ invariance
the elctromagnetic coupling is not allowed to contain Pauli terms \cite{1,2}}.
This is due to the rotation invariance of the matrix $U$ in isospace, which 
leads to the following relation:
\begin{equation}
 U^{\dag} \frac{\vec{\tau}}{2} U = \frac{1}{2}\vec{\tau}+iU^{\dag}
[\vec{\Phi}\times \frac{\partial U}{\partial \vec{\Phi}}] .
\end{equation}
It is suggestive to require the simultaneous parity and $\gamma _5$-
invariance also for all $\pi$-meson-baryon couplings. As long as one can 
neglect the mass difference of the $\Lambda$- and $\Sigma$ particle, the
pairs $\Sigma ^+$, $\frac{1}{\sqrt{2}}(\Lambda-\Sigma ^0)$ and 
$\frac{1}{\sqrt{2}}(\Lambda+\Sigma ^0)$, $\Sigma ^{-}$ can be introduced 
\cite{12}. Then the arguments given above for the $\pi$-meson-nucleon 
interaction can be carried over to all isospin doublets literally.\\

We thank Dr Symanzik for valuable suggestions.
One of us (H. R.) thanks the "G\"orres- Gesellschaft der Wissenschaften" for a
fellowship.\\

\setcounter{equation}{0}
\def\theequation{A\arabic{equation}}
\section*{Appendix}

In this appendix the asymptotic equality of the fields $\psi (x)$ and
$\psi '(x)$ shall be shown. We will require, that the function $U$ from Eqs.(3)
and (4) can be inverted and that the field $\psi '(x) = (cU)^{-1} \psi (x)$
can be expanded into powers of $f/m\Phi (x)$. The adiabatic hypothesis
($f/m \rightarrow 0$ for $t \rightarrow \pm \infty$), used in the past,
lets it appear being plausible that the two fields $\psi '(x)$ and $\psi (x)$
agree in the limit of large time.\\

To proceed more rigorously, the results of the work of Zimmermann \cite{13}
can be used. For this purpose we define ingoing and outgoing fields by the 
relation \cite{13}
\begin{equation}
\left.
\begin{array}{l}
\psi_{in/out}(x) = \psi (x)+ \int S_{ret/adv}(x-x')(\gamma _{\mu}
\frac{\partial}{\partial x_{\mu}}+M)\psi (x')d^4x' 
\\ 
\\
\psi _{in/out}'(x) = \psi '(x) + \int S_{ret/adv}(x-x')(\gamma _{\mu}
\frac{\partial}{\partial x_{\mu}}+M)\psi '(x')d^4x'.
\end{array}
\right\} 
\end{equation}

The proof of the equality of the two asymptotic fields is accomplished, if the 
two following points can be proven:\\
I. A one-nucleon state $\Psi _{\alpha}$ of the Hilbert space can be 
constructed with the help of the field $\psi _{in/out}'(x)$ as well as  with 
the help of the field $\psi _{in/out} (x)$ from the vacuum $\Omega$:
\begin{eqnarray}
\Psi _{\alpha} = \int \psi^{*}_{in/out}(x)f_{\alpha}(x)d^3x|\Omega\rangle
               = \int\psi^{*'}_{in/out}(x)f_{\alpha}(x)d^3x|\Omega\rangle 
\end{eqnarray}
In these expressions $f_{\alpha}$ is a solution of the free Dirac equation.\\
II. The field $\psi '_{in/out}(x)$ possesses with the fields 
$\psi _{in/out}(x)$ and $\Phi _{in/out}(x)$ the same commutation relations 
as the field $\psi _{in/out}(x)$ has with these field operators.\\

From points I and II it follows for all Hilbert-space states $\Psi_1$ and 
$\Psi_2$:                 
\begin{eqnarray}
\int f^{*}_{\alpha}(x)\langle \Psi_1|\psi '_{in/out}(x)|\Psi_2\rangle d^3x =
\int f^{*}_{\alpha}(x)\langle \Psi_1|\psi _{in/out}(x)|\Psi_2\rangle d^3x 
\nonumber   
\end{eqnarray}
which shows the complete equivalence of the two asymptotic fields.\\

Point I can be obtained from the Lorentz invariance of the field operators
$\psi (x)$ and $\psi '(x)$. Denoting by $\Psi _p$ a one-nucleon state with
momentum $p$ and $-p^2=M^2$, it is                        
\begin{eqnarray}
\langle \Omega|\psi_{in}(x)|\Psi_p\rangle = \frac{1}{(2\pi)^{\frac{3}{2}}} 
e^{ipx} u_p
\end{eqnarray}
and
\begin{eqnarray}
\langle \Omega|\psi '_{in}(x)|\Psi_p\rangle = 
\langle \Omega|\psi '(x)|\Psi_p\rangle = 
\frac{1}{(2\pi)^{\frac{3}{2}}} e^{ipx} u_p.
\end{eqnarray}
In order that on the right side of Eq.(A.4) the plane wave has the same
factor as in Eq.(A.3) the renormalization constant $c$ from Eq.(4),
respectively Eq.(2), must obey the equation
\begin{eqnarray}
 c= (2\pi)^{\frac{3}{2}}
\bar{u}_p \langle\Omega|U^{-1}(0)\psi(0)|\Psi _p\rangle .
\end{eqnarray}
From the fact, that according to (A.1) the ingoing and outgoing fields obey
the free Dirac equation, it moreover follows 
\begin{eqnarray} 
 \langle\Omega |\psi '_{in/out}(x)|\Psi \rangle =
\langle \Omega|\psi _{in/out}(x)|\Psi \rangle =0
\end{eqnarray}
for all states of the Hilbert space with $-p^2 \neq M^2$. From (A.3), (A.4)
and (A.6) the statement I follows.\\

Furthermore (A.3), (A.4) and (A.6), together with the corresponding relations 
for the field $\Phi _{in/out}(x)$, allow the calculation of the vacuum 
expectation values of the wanted commutation relations \cite{13}. Therefore the
latter ones are identical for $\psi _{in/out}'(x)$ and for $\psi _{in/out}(x)$
fields. It is more difficult to prove the full statement II. In order that
the equality of the commutation relations is valid in general and not only for 
the vacuum expectation values, it is necessary to show, that all commutation 
relations are c numbers. Expanding $\psi '(x)$ in a power series in $f/m$
it seems possible to perform such a proof in analogy to Zimmermann \cite{13}
for every approximation in $f/m$, if the appearing operator products are
suitably defined. 


\end{document}